\documentclass[]{aa}  

\usepackage{graphicx}
\usepackage{txfonts}
\usepackage{hyperref}
\usepackage{url}
\begin{document} 

\def\teff{${T}_{\rm eff}$}
\def\kms{{km\,s}$^{-1}$}
\def\logg{$\log g$}
\def\micro{$\xi_{\rm t}$}
\def\macro{$\zeta_{\rm RT}$}
\def\rad{$v_{\rm r}$}
\def\vsini{$v\sin i$}
\def\ebv{$E(B-V)$}
\def\kepler{\textit{Kepler}}

   \title{An all-sky catalogue of stellar reddening values}

\author{E.~Paunzen\inst{1}
        \and M.~Netopil\inst{2}
        \and M.~Pri{\v s}egen\inst{3}
                                \and N.~Faltov{\'a}\inst{1}
        }
\institute{Department of Theoretical Physics and Astrophysics, Masaryk University, Kotl\'a\v{r}sk\'a 2, 611\,37 Brno, Czechia, \email{epaunzen@physics.muni.cz}
\and Kuffner Observatory, Johann-Staud-Straße 10, A-1160 Wien, Austria
\and Advanced Technologies Research Institute, Faculty of Materials Science and Technology in Trnava, Slovak University of Technology in Bratislava, Bottova 25, 917 24 Trnava, Slovakia}

\date{}

 
  \abstract
   {When observing astronomical objects, we have to deal with extinction (i.e. the absorption and scattering of the emitted radiation by dust 
   and gas between the source and the observer). Interstellar extinction depends on the location of the object and the wavelength. The different extinction laws describing these effects are difficult to estimate for a small sample of stars.}
   {Many sophisticated and automatic methods have recently been developed for estimating astrophysical parameters (age and metallicity, for example) depending on the reddening, which is normally treated as a free parameter within the corresponding estimations. However, many reddening values for stars have been published over the last few decades, most of which include observations in the ultraviolet, which are essential for a good estimation but are essentially no longer taken into account. }
   {We searched the literature through the end of 2022 for published independent reddening values of stellar objects based on various methods that exclude estimates from reddening maps. In addition, we present new reddening estimates based on the classical photometric indices in the Geneva, Johnson, and Str{\"o}mgren-Crawford systems. These are based on well-established and reliable calibrations.}
   {After a careful identification procedure and quality assessment of the data, we calculated the mean reddening values of 157\,631 individual available measurements for 97\,826 objects. We compared our results with the ones from recent automatic pipeline values, including those from the \textit{Gaia} consortium. In addition, we chose star cluster members to compare their mean
   values with estimates for the corresponding aggregates. Within the different references, we find several statistically significant offsets and trends and discuss possible explanations for them.}
         {Our new catalogue can serve as a starting point for calibrating and testing automatic tools such as isochrone and
  spectral energy distribution fitting. Our sample covers the whole sky, including the Galactic field, star clusters, and Magellanic Clouds, and so can be used for a variety of astrophysical studies.}


   \maketitle

%
\section{Introduction} \label{introduction}

Extinction and reddening caused by interstellar dust affect the detected radiation
from most observable astronomical sources. The problem worsens when objects lie close to the
Galactic plane due to the high dust density \citep{2016MNRAS.463.3604H}. An error in determining the 
reddening will have severe consequences, for example for drawing colour-magnitude diagrams and 
determining astrophysical parameters such as the effective temperature, luminosity, mass, and age
\citep{2006MNRAS.371.1641P,2009A&A...501..941H}. 

However, \citet{1963BAN....17..115J} conducted multi-colour photometry of O- and B-type stars 
and found no unique reddening law. In addition, they concluded that there is a minor variation in the total-to-selective extinction ratio ($R_{\rm V}$) with Galactic longitude. This was later confirmed by \citet{1975ApJ...196..261S} from the spatial variation in the wavelength of maximum polarisation. Finally, \citet{1999PASP..111...63F} summarised the knowledge
about the interstellar reddening law, $R_{\rm V,}$ varying for different sight lines.
As a further complication, the interstellar extinction is variable depending on the wavelength \citep{1989ApJ...345..245C}.
The combination of these two effects leads to different extinction coefficients for
different photometric filters for a specific reddening law. This must be considered 
for the ultraviolet (UV) and infrared region \citep{1999PASP..111...63F}, which is normally not accessible from the ground.
An excellent overview of the relevant processes and methods can be found in \citet{1990ARA&A..28...37M}. 

Reddening measurements for the interstellar medium in the past relied on three techniques: (1)
star counts \citep{1923AN....219..109W}, (2) spectroscopic and photometric measurements with their calibrations
\citep{1980AandAS...42..251N}, and (3) main sequence fitting of colour-colour and
colour-magnitude diagrams \citep{1994ApJ...429..694L}. Each of these techniques has its own advantages and disadvantages.

With the availability of precise photometric data from the UV to infrared
region, spectral energy distribution (SED) fitting techniques \citep{2012ApJ...757..166B} 
are now widely used to build 3D reddening maps. This technique needs, in addition to a good knowledge
of the stellar astrophysical parameters, a set of stellar flux models that introduces another
source of uncertainty \citep{2012A&A...547A.108L}.   

The astrometric results from the \textit{Gaia} satellite \citep{2018A&A...616A...1G} open a new
window for the determination of astrophysical parameters based on
precise parallaxes. The corresponding final catalogue will include ultra-precise parallaxes
(and thus distances) and auxiliary measurements such as photometry and spectroscopy, which
makes the reddening determination a critical issue.  

In this paper we present a catalogue of mean reddening values compiled from the literature,
which, to our knowledge, does not exist elsewhere.
It provides a source of standard values for future calibration purposes, especially as a starting
point for the SED fitting procedures and new \textit{Gaia} data releases. 

\begin{figure}
\begin{center}
\includegraphics[width=\columnwidth, clip]{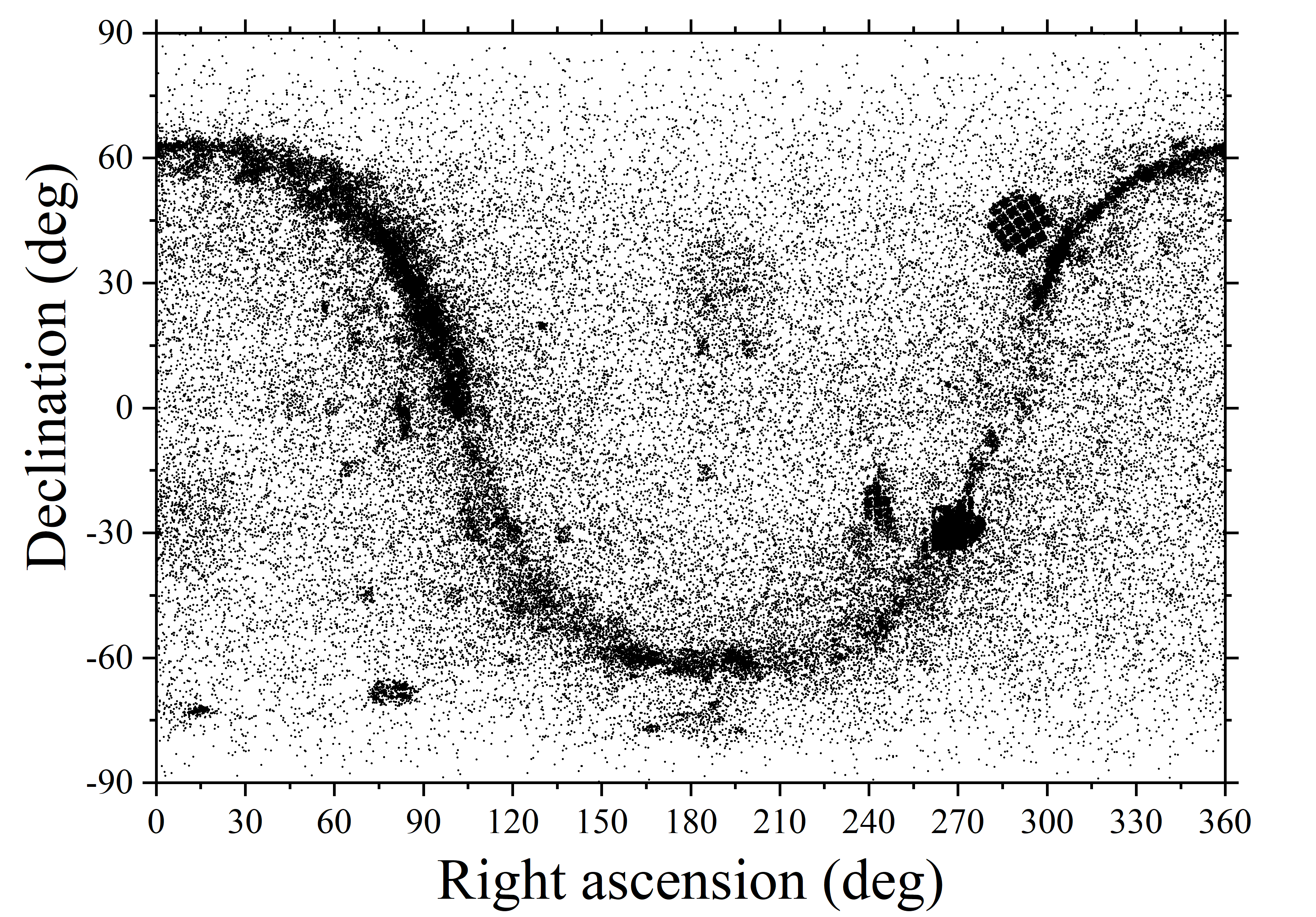}
\caption{Distribution of the catalogue stars on the sky. The Galactic disk, several star clusters,
the Magellanic Clouds, 
and the $Kepler$ field at [300$\degr$,+45$\degr$] are visible.}
\label{alpha_delta} 
\end{center} 
\end{figure}

\section{Sample selection and data sources} \label{data_sources}

Besides the inclusion of references already known to us, the literature was searched until the end of the year 2022 for reddening estimates of stellar objects based on various methods, and no time limit was set for the past. Almost all larger catalogues are already available from the CDS/VizieR, and we used some reddening related keywords of the unified content descriptor -- such as \textit{phot.color.excess} or \textit{phys.absorption} -- to identify appropriate papers. A pre-selection based on the catalogue title was made because of the many returned references. Furthermore, we do not cover works that include reddening estimates for fewer than ten objects. This certainly results in an incomplete list, but a complete census cannot be gathered anyway. The remaining references were checked to see if these include independent reddening estimates -- thus, we excluded works that adopt estimates of others that are, for example, already included in our list. The compiled references and the number of studied objects are listed in Table\,\ref{references_catalogue}. We do not cover results of reddening maps or \textit{Gaia} data, as we intend to provide a catalogue for their validation.

The reddening values are not always listed as $E(B-V)$ but also in the Str{\"o}mgren-Crawford 
\citep{1966ARA&A...4..433S} and Two Micron All Sky Survey \citep[2MASS;][]{2mass} photometric
systems. For the transformation to $E(B-V)$, we used the following transformations 
\citep{1999PASP..111...63F,2019ApJ...877..116W}:

\begin{eqnarray}
E(B-V) &=& 1.35E(b-y) = 0.36E(V-K_{\rm S}) =  \nonumber \\
&=& 2.00E(J-K_{\rm S}) = 0.32A_{\rm V} = 2.78A_{\rm K_{\rm S}}.          
\end{eqnarray}

The errors for the colour excess ratios of the different photometric bands 
listed in \citet{2019ApJ...877..116W} are typically below 0.04\,mag. We also
considered a non-standard reddening law if listed in the individual references.
In addition to the references from the literature, we also calibrated reddening values in the 
Geneva seven-colour, Johnson, and Str{\"o}mgren-Crawford photometric systems. In the following,
we overview the corresponding calibrations and data sources.

{\it The Geneva photometric sample:} The Geneva photometric catalogue \citep{2022A&A...661A..89P}
belongs to one of the most homogeneous datasets, based on a unique instrumentation 
and reduction procedure in both hemispheres \citep{Cramer1999}. The about 43\,000 measured stars in the General Catalogue of 
Photometric Data \citep[GCPD\footnote{http://gcpd.physics.muni.cz/};][]{MMH1997} cover all spectral types and luminosity classes. 
For O/B type stars, the reddening free $X/Y$ parameters \citep{Cramer1982,Cramer1999} enable an estimate of the reddening in $E[B-V]$ that can be transformed to the Johnson $UBV$ equivalent $E(B-V) = 0.842 \times E[B-V]$.
In the first step we used the parameter space $-0.3 \leq X \leq +1.4$ and $-0.07 \leq X \leq +0.15$ to extract the most likely 
O- and B-type stars population \citep[see e.g.][]{Cramer1999,Netopil2017}. The sample was then refined by a cross-match with SIMBAD and a check of available spectral classifications. This was done on the basis of coordinates (a
search radius of ten arcseconds was used) and apparent magnitudes.
The final cleaned sample includes 10\,233 stars with a reddening based on Geneva photometric data.

{\it The Johnson photometric sample:} We employed the $Q$ parameter, which is usable for early-type (i.e. hotter than B9) stars.
It utilises the colours $U-B$ and $B-V$ and their ratio \citep{Gutierrez1975}. It has been widely applied, especially for young open clusters \citep[for example,][]{Moitinho1997,Cummings2018}. For this work, the mean Johnson $UBV$ 
measurements from the GCPD were taken. As the next step, the $Q$ values for different spectral types were calculated, and all 
relevant hot objects were identified.
For this, the spectroscopic data compilation by \citet{Skiff2014} and further spectroscopic data from the SIMBAD database
were used. If no spectroscopic information was available, objects were deleted from the list rather than introducing wrong
reddening values for highly reddened cool-type objects, for example. The final list of reddening values based on the $Q$-method
includes 9878 stars.

{\it The Str{\"o}mgren-Crawford photometric sample:} The $uvby\beta$ photometric system has been widely and successfully used since its introduction almost 60 years ago \citep{Stroemgren1963}. Multiple calibrations for different stellar groups have been published since then.
We used the catalogue of mean $uvby\beta$ values by \citet{Paunzen2015}. It includes 60\,668 stars with unweighted mean 
indices and their errors. From this compilation, only stars with the complete set of $uvby\beta$ measurements were considered.
This results in a list of 32\,529 stars in total.
We used the program TEMPLOGG \citep{Kupka2001} to calculate the reddening values. 
This program is a menu-driven application written in C-Shell script. It provides convenient access to the different Fortran
codes that perform the inversions from observed colour indices to fundamental
parameters, and it also enables the user to process tables of photometric data for a
list of stars in batch form. It applies the calibrations from
\citet{Schuster1989}, \citet{Napiwotzki1993}, and \citet{Domingo1999} and covers the whole spectral type range.

The identifications in the references (Table \ref{references_catalogue}) are manifold. They 
range from commonly known IDs (for example, HD and HIP numbers) to coordinates in several 
epochs. Especially the Lausanne code numbering system (LID) used in the GCPD, is sometimes
troublesome. However, a common object identification must be done before the final means can be calculated.
Our primary source for this task is the 2MASS catalogue. It is 
deep enough for our purpose but also covers the brightest stars. All objects were carefully
checked within the 2MASS survey. If no unambiguous identification was possible, we skipped the
measurements. In total, only about 1700 data points were skipped. This is a small number compared
to the over 120\,000 measurements in the final catalogue.

In the last step, we calculated unweighted mean values and their standard deviations. Finding criteria
for introducing meaningful weights on a justified basis was impossible. As expected for many objects and references, some mean reddening values exhibit large errors up to 2\,mag. 
The reasons for that could be wrong identifications or simply very divergent measurements.

The final catalogue includes 157\,631 individual available measurements for 97\,826 objects and will be available in electronic form, an excerpt is listed in Table\,\ref{table_mean_reddening}.

\begin{table*}[t]
\caption{References used for our catalogue, with the number of measurements included. The last three photometric sources
are described in Sect. \ref{data_sources}. The references are sorted by the date of publication.}
\label{references_catalogue}
\begin{center}
\begin{tabular}{cccccc}
\hline
\hline
Reference & $N_{\rm meas}$ & Reference & $N_{\rm meas}$ & Reference & $N_{\rm meas}$ \\
\hline
\citet{1956ApJS....2..389H}     &       1257    &       \citet{2009ApJ...693L..81M}     &       30      &       \citet{2017ApJ...841...84N}     &       55      \\
\citet{1977AandAS...27..343D}   &       355     &       \citet{2010AandA...514A..59K}   &       545     &       \citet{2017ApJ...842...42M}     &       59      \\
\citet{1977AJ.....82..113S}     &       1437    &       \citet{2010AandA...521A..26P}   &       28      &       \citet{2017ApJ...848..106W}     &       34      \\
\citet{1980AandAS...40..199P}   &       8560    &       \citet{2010ApJ...708.1628M}     &       114     &       \citet{2017BlgAJ..27...10N}     &       10      \\
\citet{1980AandAS...42..251N}   &       12\,138 &       \citet{2011ApJ...726...19W}     &       241     &       \citet{2018AandA...610A..30A}   &       60      \\
\citet{1985ApJS...59..397S}     &       1414    &       \citet{2012AandA...539A.143N}   &       20      &       \citet{2018AandA...610A..64V}   &       143     \\
\citet{1993AJ....105..980M}     &       38      &       \citet{2012AandA...544A.136R}   &       150     &       \citet{2018AandA...616A.124A}   &       126     \\
\citet{1993SAAOC..15...53K}     &       256     &       \citet{2012AJ....143...99T}     &       202     &       \citet{2018AJ....155..196R}     &       1310    \\
\citet{1994AandA...290..609D}   &       184     &       \citet{2012ApJ...746..154P}     &       124     &       \citet{2019AandA...631A.124A}   &       136     \\
\citet{1994AJ....107.1577A}     &       105     &       \citet{2012ApJ...748..107P}     &       98      &       \citet{2019ApandSS.364..175C}   &       504     \\
\citet{1996PASP..108..772P}     &       3761    &       \citet{2013AandA...552A..92M}   &       114     &       \citet{2020AandA...634A..18C}   &       128     \\
\citet{1997AandA...321..236K}   &       133     &       \citet{2013AJ....145..125V}     &       29      &       \citet{2020AandA...638A..59B}   &       174     \\
\citet{1999AandA...344..263K}   &       61      &       \citet{2013MNRAS.430.2169R}     &       134     &       \citet{2020ApJ...893...56M}     &       835     \\
\citet{1999AandAS..134..525B}   &       96      &       \citet{2014AandA...562A..69K}   &       692     &       \citet{2020ApJ...905...38R}     &       814     \\
\citet{2001AJ....121.2159G}     &       232     &       \citet{2014AJ....147..137L}     &       225     &       \citet{2021AandA...645A..85M}   &       3314    \\
\citet{2002BaltA..11..441Z}     &       340     &       \citet{2015AandA...577A..23M}   &       19      &       \citet{2021AJ....161..176R}     &       3996    \\
\citet{2003AandA...404..913S}   &       825     &       \citet{2015AandA...581A..68C}   &       226     &       \citet{2021AJ....162..224S}     &       1084    \\
\citet{2003AandA...410..905P}   &       163     &       \citet{2015AJ....150...41G}     &       136     &       \citet{2021ApJS..253...22X}     &       26\,843 \\
\citet{2004AandA...418..989N}   &       16\,631 &       \citet{2015MNRAS.446..274R}     &       247     &       \citet{2021MNRAS.503.3660P}     &       101     \\
\citet{2004AandA...422..527S}   &       497     &       \citet{2015MNRAS.446.3797H}     &       144     &       \citet{2021MNRAS.508.4047R}     &       48      \\
\citet{2004ApJ...616..912V}     &       402     &       \citet{2015MNRAS.454.2863H}     &       190     &       Geneva  &       10\,233 \\
\citet{2005ApJ...623..897L}     &       105     &       \citet{2016AandA...585A.141G}   &       8056    &       UBV     &       9878    \\
\citet{2005MNRAS.358..563M}     &       49      &       \citet{2017AandA...606A..76C}   &       113     &       uvby    &       32\,529 \\
\citet{2008AJ....135..631K}     &       2411    &       \citet{2017AJ....153...16S}     &       1528    &               &               \\
\citet{2008MNRAS.389.1336K}     &       74      &       \citet{2017AJ....154...31G}     &       289     &               &               \\
\hline
\end{tabular}
\end{center}
\end{table*}

\begin{figure}
\begin{center}
\includegraphics[width=0.8\columnwidth]{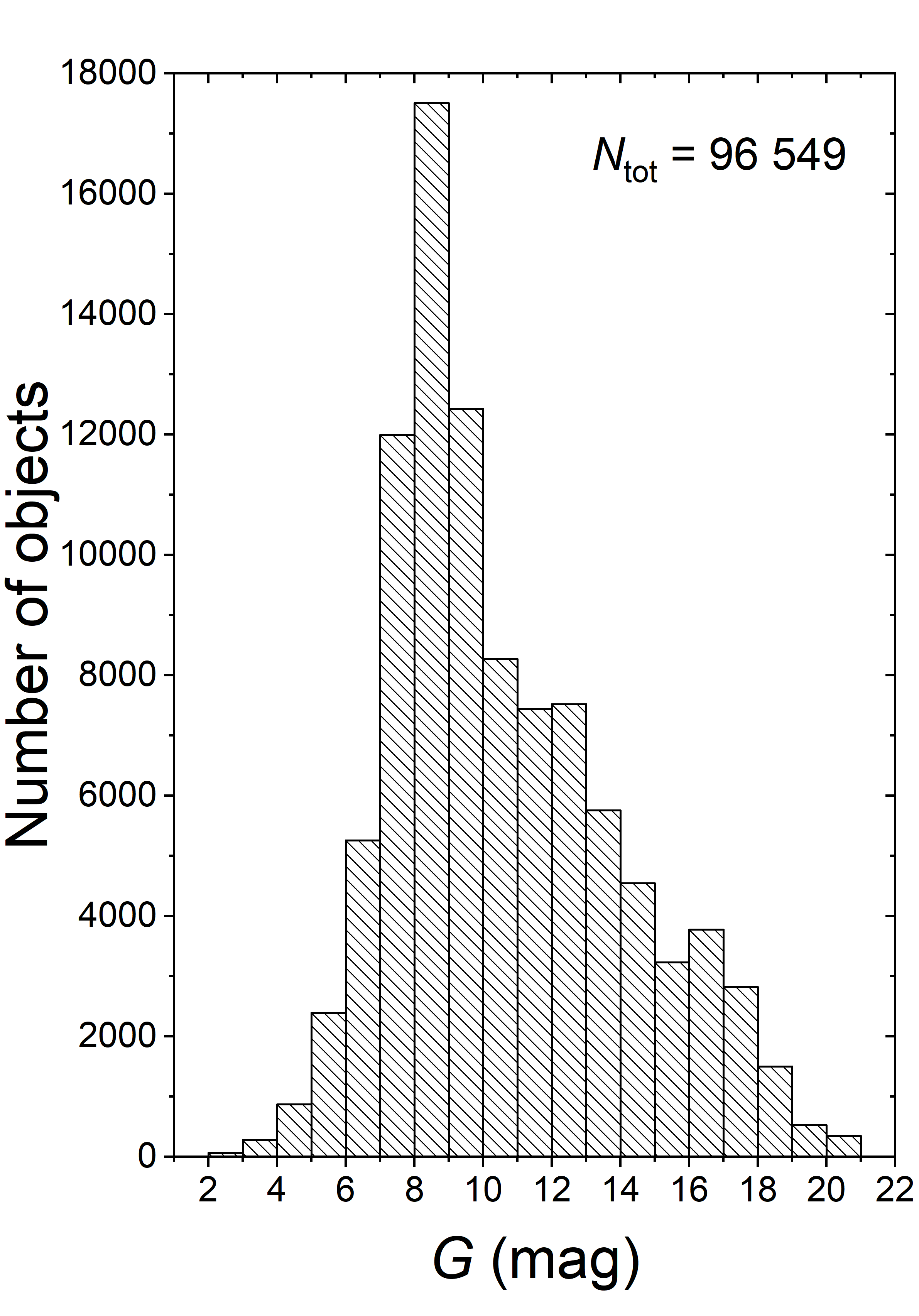}
\caption{Histogram of the $G$ magnitudes from the \textit{Gaia} DR2 for 96\,549 objects
that were successfully cross-matched. The second maximum in the distribution at about
$G \approx 16.5$\,mag is due to stars of the Galactic Bulge.}
\label{histogram_g} 
\end{center} 
\end{figure}

\section{Analysis} \label{analysis}

Our primary source for common identification is the 2MASS catalogue. For the further analysis,
we cross-matched it with the \textit{Gaia} Data Release 2 and 3 (DR2 and DR3) catalogues \citep[][]{2018A&A...616A...1G,2021A&A...649A...1G} using the coordinates
and $J$ as well as $G$ magnitudes. Again, we were conservative, rather than excluding a 
\textit{Gaia} source instead of including a false identification. We could identify 96\,549 from
the 97\,826 objects (or 98.69\%) of our catalogue.

The final catalogue represents a sample of stars distributed over the whole sky, as shown in Fig. \ref{alpha_delta}.
Several important Galactic features, such as the disk, the central and Bulge region, and the north
and south poles, are seen. However, the Magellanic Clouds are also visible.

In Fig. \ref{histogram_g} the histogram of the $G$ magnitudes is shown. Our sample includes objects with magnitudes of 1.7 to 21.2,  with maxima at 8.5 and 16.5 magnitudes. The 16.5 magnitude maximum is caused
by objects from the Galactic Bulge \citep{2008IAUS..245..323M}. It shows a statistically
significant number of stars in all magnitude bins, which can be used for calibration.

\subsection{The references for assessing the results} \label{references_assesment}

We selected six recent and widely used sources
of reddening values to compare and analyse our values with the literature: \citet{2018A&A...616A...8A}, \citet{2019ApJ...887...93G}, \citet{2019A&A...628A..94A}, \citet{2019MNRAS.483.4277C}, \citet{2021A&A...649A...1G}, and \citet{2022A&A...658A..91A}.
From these, the work by \citet{2019ApJ...887...93G} presents a 3D map, whereas the other three
sources list reddening values for individual stars.

We included a comparison with results from the \textit{Gaia} DR2 \citep{2018A&A...616A...8A} because many papers devoted to star clusters and stellar astrophysics are based on it. Most of them use reddening values for the analysis.

For all six references, we have to emphasise that no filter in the UV region (bluewards of 400\,nm)
was used for the reddening estimates. It is well known that indices such as Johnson $(U-B)$ and 
Str{\"o}mgren $(u-b)$ are most efficient for such a purpose. This is a severe drawback. We note that
the $G$ and $G_{\rm BP}$ filters of the \textit{Gaia} photometric system measure the flux down to about 330\,nm, with a minimal efficiency compared to the other, redder regions. Furthermore, the filters are much too broad (blue cutoffs at
680 and 1000\,nm, respectively) to be useful for the reddening estimate.

We also notice that the StarHorse \citep{2019A&A...628A..94A} and 
StarHorse2021 \citep{2022A&A...658A..91A} compilations include negative reddening values. It is
written in the papers that these values must be treated with caution, but no other comments
for the reasons are given.

\begin{table*}
\caption{Statistical parameters of the differences for the references listed below.} 
\footnotesize
\label{statistics_published_catalogues}
\begin{center}
\begin{tabular}{cccccccccccc}
\hline
\hline
        &       (2)$-$(3)       &       (2)$-$(4)       &       (2)$-$(5)       &       (2)$-$(6)       &       (2)$-$(7)       &       (3)$-$(4)       &       (3)$-$(5)       &       (3)$-$(6)       &       (3)$-$(7)       &       (4)$-$(5)       &       (4)$-$(6)       \\
\hline
Mean    &       $-$0.098        &       $-$0.117        &       $-$0.021        &       $-$0.074        &       $-$0.055        &       +0.078  &       +0.105  &       +0.078  &       +0.100  &       +0.033  &       +0.009 \\
Median  &    $-$0.125   &       $-$0.090        &       $-$0.009        &       $-$0.069        &       $-$0.041        &       +0.059  &       +0.095  &       +0.062  &       +0.095  &       +0.001  &       $-$0.009 \\
STD     &       +0.211  &       +0.216  &       +0.163  &       +0.142  &       +0.191  &       +0.228  &       +0.198  &       +0.188  &       +0.232  &       +0.199  &       +0.148 \\
Skewness        &       +1.779  &       $-$0.165        &       +0.953  &       +0.046  &       $-$0.627        &       $-$2.099        &       $-$1.407        &       $-$1.759        &       $-$1.964        &       +1.208  &       +0.160 \\
Kurtosis        &       +6.929  &       +0.945  &       +8.553  &       +1.332  &       +9.798  &       +12.204 &       +9.913  &       +10.867 &       +11.018 &       +8.145  &       +11.200 \\
$N$     &       21\,897 &       22\,183 &       35\,736 &       33\,077 &       25\,613 &       35\,538 &       49\,478 &       44\,665 &       35\,333 &       57\,646 &       54\,173 \\

\hline
        &       (4)$-$(7)       &       (5)$-$(6)       &       (5)$-$(7)       &       (6)$-$(7)       &               &       (1)$-$(2)       &       (1)$-$(3)       &       (1)$-$(4)       &       (1)$-$(5)       &       (1)$-$(6)       &       (1)$-$(7)       \\
Mean    &       +0.060  &       $-$0.023        &       +0.002  &       +0.031 &               &               +0.167  &       $-$0.033        &       +0.076  &       +0.114  &       +0.093  &       +0.139  \\
Median  &       +0.018  &       $-$0.009        &       +0.028  &       +0.034  &       &               +0.134  &       $-$0.048        &       +0.001  &       +0.026  &       +0.006  &       +0.026  \\
STD     &       +0.224  &       0.132   &       +0.199  &       +0.124  &               &    +0.278       &       +0.320  &       +0.477  &       +0.365  &       +0.369  &       +0.491  \\
Skewness        &       +1.399  &       $-$1.380        &       $-$1.845        &       $-$1.289   &             &       +2.387  &       +5.912  &       +3.828  &       +3.942  &       +4.680  &       +4.058  \\
Kurtosis        &       +23.112 &       +14.419 &       +16.400 &       +26.513 &               &               +17.992 &    +105.390     &       +27.052 &       +26.673 &       +33.818 &       +29.787 \\
$N$     &       47\,227 &       74\,708 &       60\,433 &       56\,451 &               &               37\,045 &51\,356        &       62\,515 &       85\,360 &       76\,725 &       67\,437 \\

\hline
\multicolumn{12}{l}{Notes:
(1) This work. (2) \citet{2019MNRAS.483.4277C}. (3) \textit{Gaia} DR2: \citet{2018A&A...616A...8A}.
(4) \textit{Gaia} DR3: \citet{2021A&A...649A...1G}. }\\
\multicolumn{12}{l}{
(5): StarHorse: \citet{2019A&A...628A..94A}.
(6) StarHorse2021: \citet{2022A&A...658A..91A}. (7) Bayestar2019: \citet{2019ApJ...887...93G}.} \\ 
\multicolumn{12}{l}{
In the Appendix, the histograms are shown in Figs. \ref{differences_1} to
\ref{differences_4}.}
\end{tabular}
\end{center}
\end{table*}

{\it \citet{2019MNRAS.483.4277C}}: They presented a 3D interstellar dust reddening map of the Galactic plane
based on \textit{Gaia} DR2, the Two Micron All Sky Survey (2MASS), and the Wide-Field Infrared Survey Explorer \citep[WISE,][]{2010AJ....140.1868W} photometry. The filters range 
from 400\,nm to 2400\,nm.
It covers the whole
Galactic longitude range with Galactic latitudes $| b | < 10\degr$. They applied a machine
learning algorithm called random forest regression to derive $E(G-K_{\rm S})$, 
$E(G_{\rm BP}-G_{\rm RP})$, and $E(H-K_{\rm S})$. In doing so, they built an empirical
training sample of stars selected from several large-scale spectroscopic
surveys, including the APO Galactic Evolution Experiment \citep[APOGEE;][]{2017AJ....154...94M}
Large Sky Area Multi-Object Fiber Spectroscopic Telescope \citep[LAMOST;][]{lamost}, and
Sloan Extension for Galactic Understanding and Exploration \citep[SEGUE;][]{2009AJ....137.4377Y} surveys.
Together with the photometry, reddening values were derived with
the star-pair technique. They state that a comparison with results in the literature
shows good agreement and that their results have typical uncertainties of about 0.07\,mag
in $E(B-V)$. We used the three listed reddening values for the individual \textit{Gaia} DR2 stars and
computed a mean reddening and its standard deviation using the following relations 
\citep{2019ApJ...877..116W}:
\begin{equation}
E(B-V) = 0.45E(G-K_{\rm S}) = 6.09E(H-K_{\rm S})                
.\end{equation}
There are 37\,045 stars that are also on our target list. 

{\it \textit{Gaia} DR2 catalogue \citep{2018A&A...616A...8A}}: It includes both the $G$-band extinction,
$A_{\rm G}$, and the $E(G_{\rm BP}-G_{\rm RP})$ reddening. They were first derived using the parallaxes and photometric 
data to get each passband's effective temperature, absolute magnitudes, and reddening
values. Then, an interpolation routine within the PARSEC evolutionary models \citep{2012MNRAS.427..127B}
for solar metallicity ($Z_\odot = 0.0152$), and a standard reddening law was applied. The authors state that
the extinction estimates are inaccurate on a star-by-star level but mostly unbiased
and so are applicable at the ensemble level.  
We transformed the reddening of $(G_{\rm BP}-G_{\rm RP})$ according to \citet{2019ApJ...877..116W}:
\begin{equation}
E(B-V) = 0.77E(G_{\rm BP}-G_{\rm RP})   
.\end{equation}
The standard deviation was derived by the given upper and lower limits for $E(G_{\rm BP}-G_{\rm RP})$. 
In total, we have 51\,356 stars in common.

{\it \textit{Gaia} DR3 catalogue \citep{2021A&A...649A...1G}}: In addition to more precise astrometric parameters, 
this data release also includes low-resolution BP/RP spectra for 219 million sources. They cover the wavelength range 330 to 1050\,nm with a resolution between 13 and 85. This resolution does not permit the measurement of individual spectral lines but can be used similar to spectrophotometry. The overall method
has not changed \citep{2022arXiv220605864C}. We note that the filter curves have been altered
compared to the \textit{Gaia} DR2. With these new data sources, the individual
reddening estimates should be expected to be closer to the already published ones, as compiled in this work.

\begin{figure*}
\begin{center}
\includegraphics[width=\textwidth]{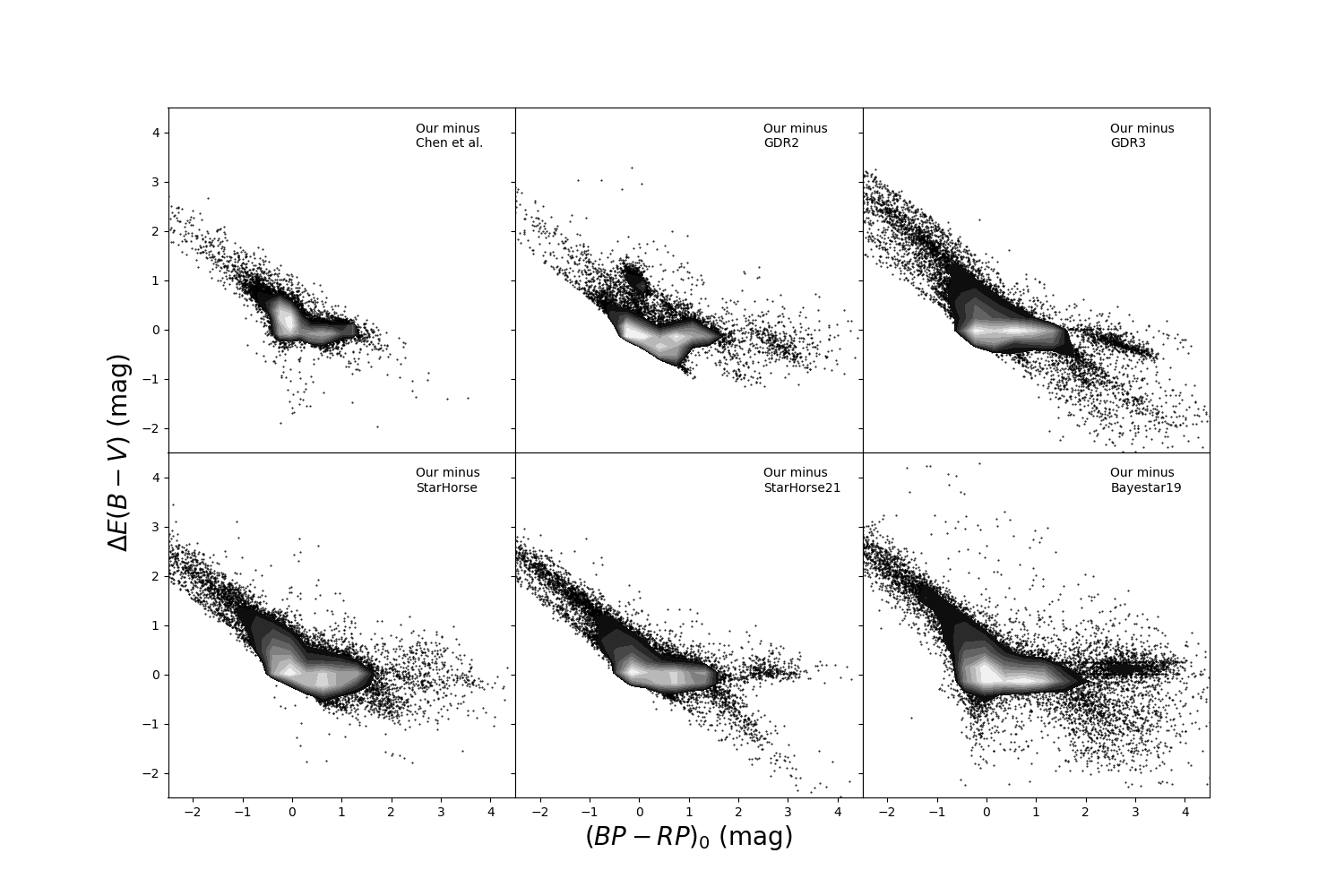}
\caption{Differences between our reddening estimates and those from the literature, as described in Table \ref{statistics_published_catalogues}.
We used contours in over-dense regions to clearly show the correlations.} 
\label{scatter_contour} 
\end{center} 
\end{figure*}

{\it StarHorse \citep{2019A&A...628A..94A}}: They combined parallaxes and photometry from the \textit{Gaia} DR2 together 
with the photometric catalogues
of Pan-STARRS 1, 2MASS, and AllWISE \citep{2013yCat.2328....0C} in order to derive
Bayesian stellar parameters, distances, and extinctions. For this purpose, the StarHorse code \citep{2018MNRAS.476.2556Q} was
applied. It is a Bayesian
parameter estimation code that compares many observed
quantities to stellar evolutionary
models. Given the set of observations plus several priors, it finds the posterior probability over a
grid of stellar models, distances, and extinctions. A mean precision of 0.20\,mag for $A_{\rm V}$ was achieved.
Our analysis used the listed line-of-sight extinction 50th percentile for 67\,437 stars in common.

{\it StarHorse2021 \citep{2022A&A...658A..91A}}: For this version, they significantly updated the
StarHorse algorithm in the context of the \textit{Gaia} DR3. Furthermore, the photometric data from 
SkyMapper DR2 \citet{2019PASA...36...33O} without the $u$ filter were included. It was
concluded that the systematic
errors of the astrophysical parameters are smaller than the nominal uncertainties for most objects.

{\it Bayestar2019 \citep{2019ApJ...887...93G}}: This is a 3D map of interstellar dust 
reddening that covers three-quarters of
the sky (i.e. declinations of $\delta > -30\degr$). It is based on the 
Pan-STARRS 1
and
2MASS colours ranging from 400\,nm to 2400\,nm. Including the astrometric data from the
\textit{Gaia} significantly improved the accuracy compared to the prior version of the 
maps \citep{2018MNRAS.478..651G}.
They group stars into small angular 
patches in the sky. Based on the available photometric measurements of each
star, its spectral type and the parallax from \textit{Gaia} (if available), they compute a probability distribution over the star's
distance and foreground dust column. Each star puts a constraint
on the line-of-sight distance versus dust column relation.
Assuming a constant reddening law, a significant number of stars
is needed along a single sight line  
to put a strong constraint on the dust density as a function
of distance. We used the distances from \citet{2021AJ....161..147B} as well as Galactic
coordinates
to estimate mean reddening values. 

\subsection{Comparison with the references} \label{general_analysis}

We statistically analysed the differences in the reddening
values of the individual references among each other and our catalogue.
We expect a normal distribution centred at zero with a certain width
or standard deviation. Two other important quantities are the third and fourth
standardised moments, the skewness and kurtosis \citep{rice2006mathematical}. 

\begin{figure}
\begin{center}
\includegraphics[width=\columnwidth]{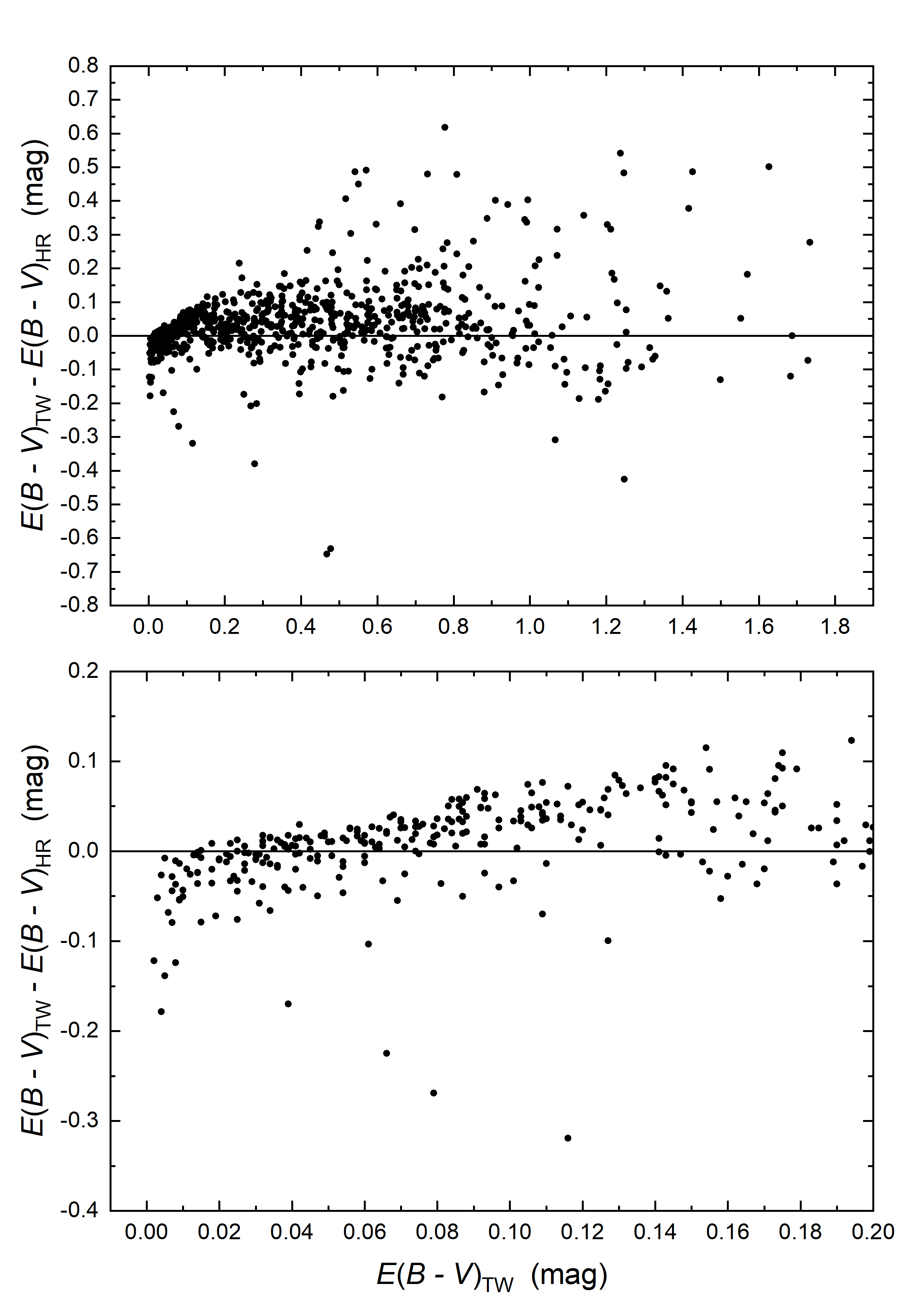}
\caption{Mean reddening values for members of 786 star clusters from our catalogue (suffix `TW') in comparison with the values
from \citet{2023A&A...673A.114H}. There is a clear trend for $E(B-V)$\,$<$\,0.2\,mag.} 
\label{comparison_HR} 
\end{center} 
\end{figure}

The skewness is a measure of symmetry
around the zero point or the lack of symmetry. In the ideal case,
the skewness for a normal distribution is zero (mean is equal to median), and any symmetric data should have a value near zero. 
Negative values indicate data that are skewed left in the sense that the left tail of the 
distribution is long relative to the right one. Values between $-$0.5 and +0.5 for a large
sample, as ours can be considered symmetric. 
The kurtosis measures whether the data are tailed relative to a normal distribution. The value for an
ideal normal distribution is three. Excess kurtosis is often used, subtracting a value of three. 

Table \ref{statistics_published_catalogues} presents all statistical values for the
different references, whereas Fig. \ref{scatter_contour} shows the data graphically. 
The histograms are shown in Figs. \ref{differences_1} to
\ref{differences_4}.
It has to be kept in mind that these characteristics are for
the extinction in $(B-V)$.

All the investigated datasets' mean and median values are compiled 
around zero. There are no apparent outliers. 
The standard deviations for the datasets of the literature vary between 0.12 and 0.23\,mag. 
These values also seem to limit the global statistical accuracy for any of these references.
The comparison with our datasets yields slightly higher standard deviations (up to 0.3\,mag).
We think that the reason is the use of UV data within our compilation. As stated before, it is
essential to include this wavelength region, especially for early-type stars \citep{1985ApJS...59..397S}.
Looking at the skewness and kurtosis values, we find that only two datasets \citep{2019MNRAS.483.4277C,2022A&A...658A..91A}
qualify as being close to normal distributions. All others deviate significantly,
especially when it comes to kurtosis. We
also calculated all values for those 8500 objects common in all datasets to exclude selection effects. The results
are the same, with slightly lower standard deviations. 

Figure \ref{scatter_contour} presents more detailed structures of the individual differences with our catalogue.
The asymmetry of the different distributions is visible, especially for the upper main sequence stars (which are, in general, at
more considerable distances from the Sun). For the cool-type stars, there is, in four cases, a similar effect in the other
direction evident. 

In general, we advise the users of the published reddening values to search all
available sources and to compare them. The object's location in the Milky Way needs to be checked for significant deviating values. 
De-reddening procedures using reddening-free indices, for instance those described in Sect.
\ref{data_sources}, are always preferable.

\subsection{Reddening values of open clusters} \label{ocls_assesment}

As the next step, we investigated the members of star clusters in our sample. The releases of the \textit{Gaia} data brought a flood of newly
discovered star clusters based on astrometric and photometric data. However, wrongly identified or already known aggregates make a
homogeneous analysis difficult. 
We used the star cluster membership lists from the catalogue of \citet{2023A&A...673A.114H} to identify possible open cluster members among the studied stars.
This catalogue contains the parameters (age, reddening, and distance) of 7167 star clusters, with more than 700 newly discovered high-confidence star clusters. In addition to this,
they also include a list of cluster members with membership probabilities for each star. For example, they performed several cross-checks with already published
catalogues and found offsets for the extinction (see Fig. 7 therein). We must emphasise that determining the cluster parameters is 
still challenging, although, from the \textit{Gaia} datasets, we already get a reasonable estimate of the distances \citep{2015A&A...582A..19N,2021MNRAS.504..356D}.
We cross-matched our sample with the lists by \citet{2023A&A...673A.114H} and identified
4270 high-confidence ($P>0.7$) star cluster members. For the comparison, we only took star clusters for which more than two members are included in
our sample, leaving us with 786 aggregates. In Fig. \ref{comparison_HR}, we present the comparison result with the published values by 
\citet{2023A&A...673A.114H}. Plotting star clusters with more than five members does not change the results.
The skewness and
kurtosis of the corresponding distribution are $-$1.99 and +8.50.
We see a wide spread of differences with a clear trend for $E(B-V)$\,$<$\,0.2\,mag. This means that their Bayesian neural network technique generally 
overestimates the reddening. 
This might be a result of the applied binning by \citet{2023A&A...673A.114H} of 
0.11\,mag in ($G_{\rm BP}-G_{\rm RP}$).
For larger reddening values, the effect reverts. 
This conclusion is also valid using mean reddening values from the six sources
we used for our comparison.
Determining the cluster parameters is still very much limited by the choice of reddening. At the same time,
the distance can be accurately estimated using the \textit{Gaia} data (which will improve further with forthcoming releases). We also now have very good
metallicities from the \textit{Gaia}-ESO \citep{2022A&A...666A.121R} and Galactic Archaeology with Hermes \citep[GALAH,][]{2021MNRAS.506..150B} surveys, for example, which could serve
as a starting point for fitting isochrones to get the age. But still, the fitting techniques need to consider all these starting values
to derive homogeneous cluster parameters \citep{2010A&A...514A..81P,2021scgr.confE..39P}.

\begin{table*}
\caption{Final catalogue (extract). The full catalogue includes 157\,631 individual available measurements for 97\,826 objects. The full table will be provided electronically in the VizieR (CDS) database.}
\label{table_mean_reddening}
\tiny
\begin{tabular}{ccccccccccccc}
\hline
\hline
ID 2MASS & $\alpha$ (2000) & $\delta$ (2000) & $G$ & (1) & std. & $N$ & (2) & (3) & (4) & (5) & (6) & (7) \\
& (deg) & (deg) & (mag) & (mag) & (mag) &  & (mag) & (mag) & (mag) & (mag) & (mag) & (mag) \\
\hline
J00000118+3851334       &       0.0050580       &       +38.8592695     &       6.5960  &       0.145   &       0.000   &       1       &               &       0.120   &               &               &               &       0.088   \\
J00000208-5153367       &       0.0090584       &       $-$51.8935374   &       7.9849  &       0.015   &       0.012   &       3       &               &               &       0.005   &       0.120   &       0.040   &               \\
J00000410+3411189       &       0.0158291       &       +34.1883085     &       8.3483  &       0.011   &       0.000   &       1       &               &       0.023   &       0.000   &       0.082   &       0.057   &       0.000   \\
J00000529+2002100       &       0.0210354       &       +20.0352750     &       9.3237  &       0.001   &       0.000   &       1       &               &       0.012   &       0.047   &       0.070   &       0.093   &       0.000   \\
J00000914+3651596       &       0.0381670       &       +36.8665193     &       9.3223  &       0.028   &       0.000   &       1       &               &       0.267   &       0.030   &       0.076   &       0.050   &       0.057   \\
J00001073+6218489       &       0.0447636       &       +62.3135773     &       16.2573 &       0.860   &       0.000   &       1       &       0.387   &               &       0.564   &       0.376   &       0.401   &       0.318   \\
J00001191+6208079       &       0.0496096       &       +62.1355781     &       11.1940 &       0.666   &       0.000   &       1       &               &               &       0.704   &       0.499   &       0.524   &       0.556   \\
J00001224-4011326       &       0.0507951       &       $-$40.1925150   &       8.0397  &       0.010   &       0.000   &       1       &               &       0.241   &       0.016   &       0.044   &       0.019   &               \\
J00001427+5049069       &       0.0594288       &       +50.8186006     &       11.6415 &       0.130   &       0.000   &       1       &               &               &       0.140   &       -0.012  &       0.100   &       0.102   \\
J00001512+2331451       &       0.0632161       &       +23.5291293     &       8.3940  &       0.001   &       0.000   &       2       &               &       0.200   &       0.000   &       0.080   &       0.029   &       0.000   \\
J00001685+3454551       &       0.0703082       &       +34.9153536     &       8.7068  &       0.085   &       0.000   &       1       &               &               &               &       0.178   &       0.064   &       0.049   \\
J00001789+1318441       &       0.0747991       &       +13.3122745     &       7.4636  &       0.003   &       0.000   &       1       &               &               &       0.005   &       0.100   &       0.073   &       0.000   \\
J00001840+2218029       &       0.0767102       &       +22.3007698     &       14.2196 &       0.070   &       0.000   &       1       &               &               &               &       -0.636  &               &       0.075   \\
J00002014-3923552       &       0.0840095       &       $-$39.3986496   &       10.2660 &       0.001   &       0.000   &       1       &               &               &               &       -0.401  &       -0.039  &               \\
J00002379-1027446       &       0.0991587       &       $-$10.4622604   &       8.0049  &       0.001   &       0.000   &       1       &               &               &       0.001   &       0.125   &       -0.005  &       0.000   \\
J00002388+2655053       &       0.0997984       &       +26.9178808     &       6.3144  &       0.022   &       0.007   &       3       &               &               &       0.000   &       0.188   &       0.005   &       0.000   \\
J00002514+6243320       &       0.1048484       &       +62.7255690     &       9.8656  &       0.298   &       0.000   &       1       &       0.067   &       0.230   &       0.261   &       0.083   &       0.154   &       0.239   \\
J00002562+6239460       &       0.1068235       &       +62.6628231     &       11.4341 &       0.458   &       0.000   &       1       &       0.310   &       0.529   &               &       0.439   &       0.361   &       0.270   \\
J00002618+6321391       &       0.1091272       &       +63.3608209     &       17.5838 &       1.469   &       0.000   &       1       &               &               &       0.146   &       0.328   &       0.681   &       0.575   \\
J00002692-1641492       &       0.1130407       &       $-$16.6971400   &       7.3668  &       0.019   &       0.000   &       1       &               &       0.418   &       0.000   &       0.184   &       0.034   &       0.000   \\
J00002716-7903431       &       0.1168140       &       $-$79.0622495   &       8.4671  &       0.001   &       0.000   &       1       &               &       0.031   &       0.001   &       0.045   &       0.053   &               \\
J00002827+6712547       &       0.1177847       &       +67.2151660     &       10.3382 &       1.506   &       0.000   &       1       &       1.240   &       0.334   &               &       1.476   &       1.468   &       1.517   \\
J00002927+6713006       &       0.1219400       &       +67.2167763     &       10.1926 &       1.580   &       0.000   &       1       &       1.007   &       0.338   &               &       1.239   &       0.864   &       1.518   \\
J00003010+2550412       &       0.1255467       &       +25.8447195     &       8.1243  &       0.015   &       0.000   &       1       &               &       0.531   &       0.015   &       0.186   &       -0.007  &       0.000   \\
\hline
\multicolumn{13}{l}{Notes: The IDs from the 2MASS survey, the coordinates and $G$ magnitudes from the \textit{Gaia} DR3, the mean reddening values from this paper} \\
\multicolumn{13}{l}{(1) based on N measurements and the values from the literature as
(2) \citet{2019MNRAS.483.4277C}. (3) \textit{Gaia} DR2: \citet{2018A&A...616A...8A}.} \\
\multicolumn{13}{l}{(4) \textit{Gaia} DR3: \citet{2021A&A...649A...1G}. (5): StarHorse: \citet{2019A&A...628A..94A}.
(6) StarHorse2021: \citet{2022A&A...658A..91A}.} \\
\multicolumn{13}{l}{(7) Bayestar2019: \citet{2019ApJ...887...93G}.}
\end{tabular}
\end{table*}

\subsection{Variable extinction of stellar objects} \label{variable_extinction}

It has been well documented that several star groups show variable extinction. Most of
these objects, like the classical T Tauri and UX Ori stars show this effect because of a
varying circumstellar environment \citep{2016AstL...42..314G,2019Ap.....62...41G}.

On a larger scale, such effects have also been observed for \ion{H}{ii} regions 
and the Galactic centre \citep{1979AJ.....84..324L,2013AAS...22114507V}. In general, it
can be said that reddening varies in regions with rapid variations in dust and gas content.

Intrinsically variable stars are at the centre of many scientific studies \citep[e.g.][]{2019A&A...623A.110G}. The periods, amplitudes, and light curve characteristics \citep{1996lcvs.book.....S} are as manifold as the underlying physical mechanisms \citep{2007uvs..book.....P}. For example, Cepheid variables and their period-luminosity-relations \citep{1912HarCi.173....1L} have allowed us to start constructing a distance ladder, which helped us explore large regions of the Universe. For such studies, the knowledge of a precise extinction
value is essential. Therefore, our catalogue can help identify possible outliers and offsets
due to incorrect reddening values.

\section{Conclusions} \label{conclusions}

Compiling mean reddening values of stellar objects is essential for calibrating and testing observations
and models. The interstellar and inter-cluster absorption must be considered for various aspects of astrophysical
studies. Its wavelength dependence and different reddening laws due to multiple mixtures of interstellar
gas and dust make it necessary to work with a statistically sound sample of objects. We present a catalogue
of mean reddening values for 97\,826 objects distributed across the visible parts of the Milky Way and the Magellanic
Clouds.

A comparison with published reddening values derived via sophisticated automatic fitting procedures reveals
systematic offsets for both Galactic field objects and star clusters. Therefore, estimating the reddening using well-established techniques and datasets, especially observations
in the UV region, is still essential. Future satellite missions such as the Quick Ultra-VIolet Kilonova surveyor \citep[QUVIK;][]{2022SPIE12181E..0BW}
will help improve this situation significantly. Also, a (re-)analysis of photometric
UV observations available in scientific archives is very much needed \citep{2014Ap&SS.354...89B}.

\begin{acknowledgements}
We are grateful to the referee for the constructive input.
This work was supported by the grant GA{\v C}R 23-07605S and the European Regional Development Fund, project No. ITMS2014+: 313011W085 (MP).
This research has made use of the SIMBAD database,
operated at CDS, Strasbourg, France and of the Two Micron All Sky Survey (2MASS), 
which is a joint project of the University of Massachusetts and the Infrared 
Processing and Analysis Center/California Institute of Technology, funded by 
the National Aeronautics and Space Administration and the National Science Foundation.
This work presents results from the European Space Agency (ESA) space mission 
\textit{Gaia}. \textit{Gaia} data are being processed by the \textit{Gaia} Data Processing and Analysis 
Consortium (DPAC). Funding for the DPAC is provided by national institutions, 
in particular the institutions participating in the \textit{Gaia} MultiLateral Agreement 
(MLA). The \textit{Gaia} mission website is https://www.cosmos.esa.int/gaia. 
The \textit{Gaia} archive website is https://archives.esac.esa.int/gaia. 
\end{acknowledgements}

\bibliographystyle{aa}
\bibliography{aa47768-23}

\begin{appendix}
\section{Histograms of the reddening differences}

\begin{figure*}
\begin{center}
\includegraphics[width=0.85\textwidth]{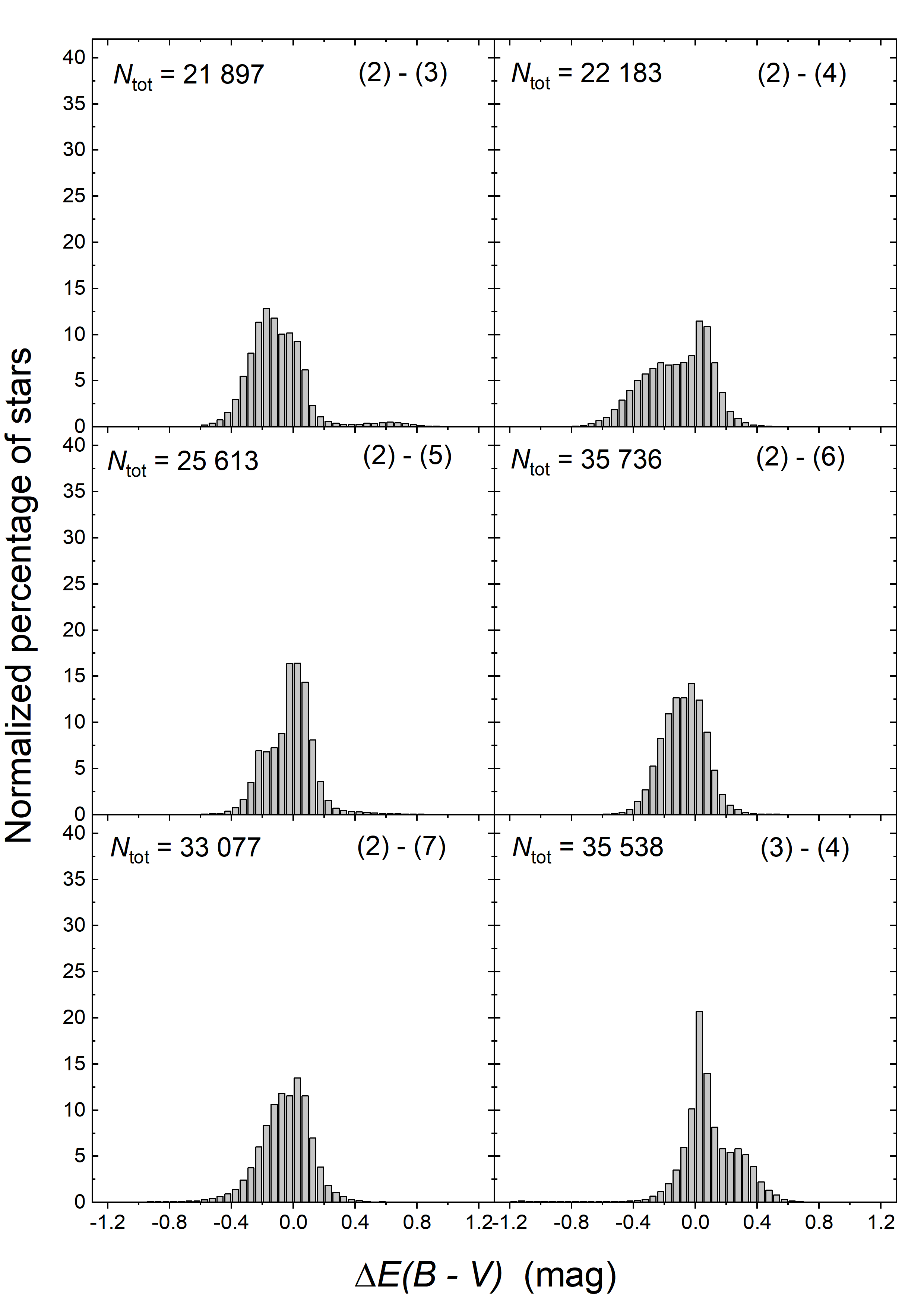}
\caption{Histograms of the differences for
(1) this work, (2) \citet{2019MNRAS.483.4277C}, (3) \textit{Gaia} DR2 \citep{2018A&A...616A...8A},
(4) \textit{Gaia} DR3 \citep{2021A&A...649A...1G}, (5) StarHorse \citep{2019A&A...628A..94A},
(6) StarHorse2021 \citep{2022A&A...658A..91A}, and (7) Bayestar2019 \citep{2019ApJ...887...93G},
as listed in Table \ref{statistics_published_catalogues}.} 
\label{differences_1} 
\end{center} 
\end{figure*}

\begin{figure*}
\begin{center}
\includegraphics[width=0.85\textwidth]{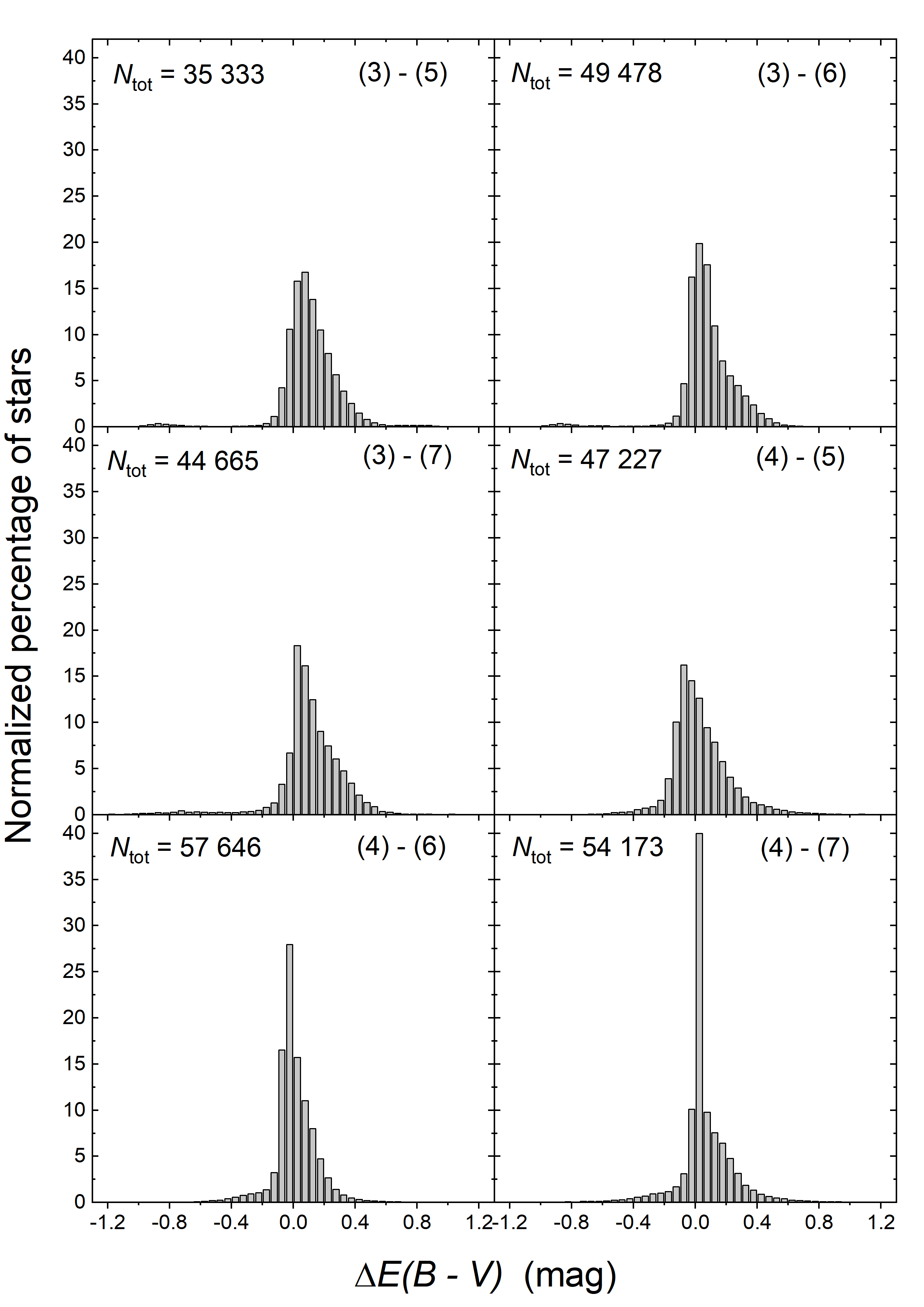}
\caption{Histograms of the differences for (1) this work, (2) \citet{2019MNRAS.483.4277C}, (3) \textit{Gaia} DR2 \citep{2018A&A...616A...8A},
(4) \textit{Gaia} DR3 \citep{2021A&A...649A...1G}, (5) StarHorse \citep{2019A&A...628A..94A},
(6) StarHorse2021 \citep{2022A&A...658A..91A}, and (7) Bayestar2019 \citep{2019ApJ...887...93G},
as listed in Table \ref{statistics_published_catalogues}.} 
\label{differences_2} 
\end{center} 
\end{figure*}

\begin{figure*}
\begin{center}
\includegraphics[width=0.85\textwidth]{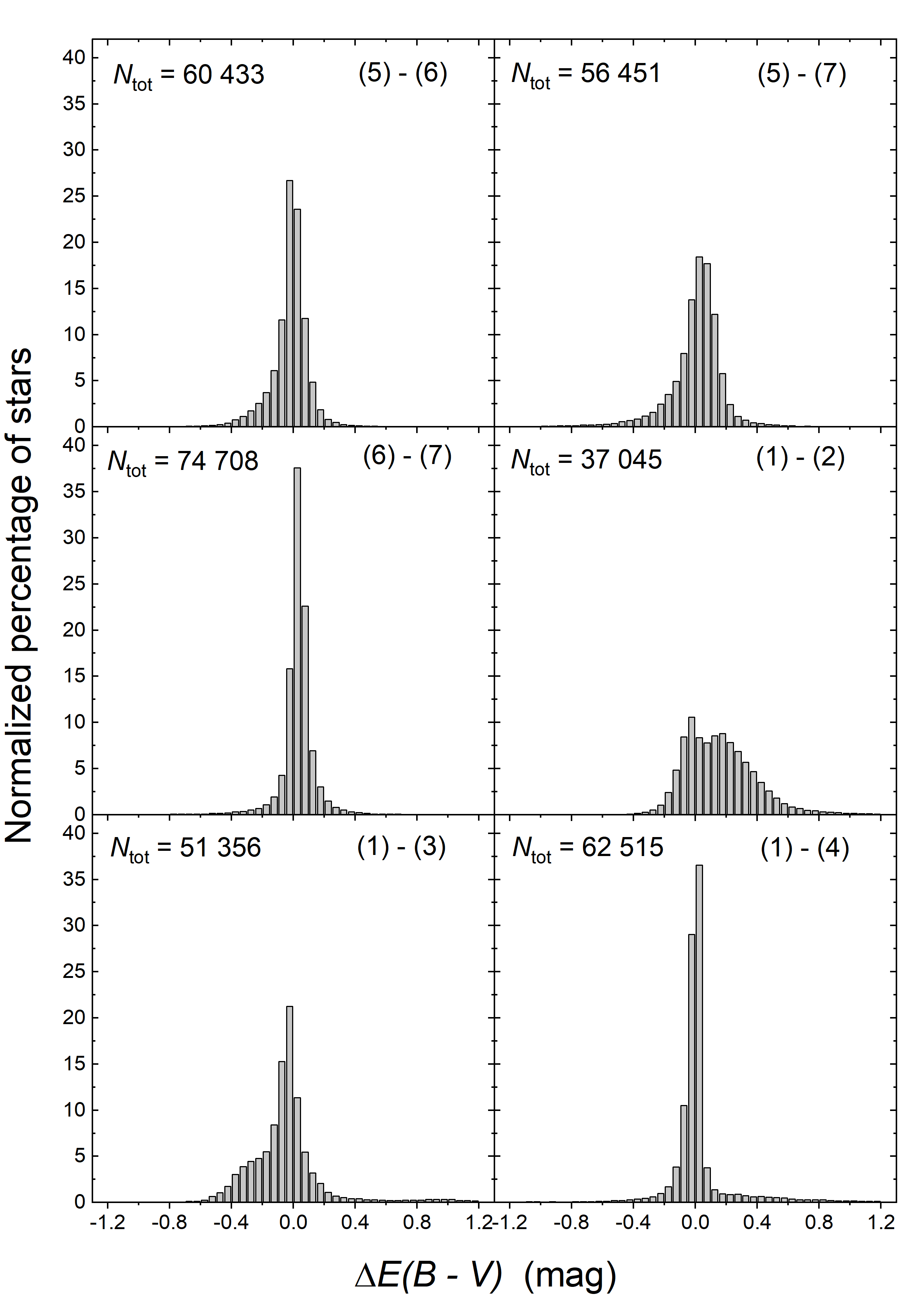}
\caption{Histograms of the differences for 
(1) this work, (2) \citet{2019MNRAS.483.4277C}, (3) \textit{Gaia} DR2 \citep{2018A&A...616A...8A},
(4) \textit{Gaia} DR3 \citep{2021A&A...649A...1G}, (5) StarHorse \citep{2019A&A...628A..94A},
(6) StarHorse2021 \citep{2022A&A...658A..91A}, and (7) Bayestar2019 \citep{2019ApJ...887...93G},
as listed in Table \ref{statistics_published_catalogues}.} 
\label{differences_3} 
\end{center} 
\end{figure*}

\begin{figure*}
\begin{center}
\includegraphics[width=0.85\textwidth]{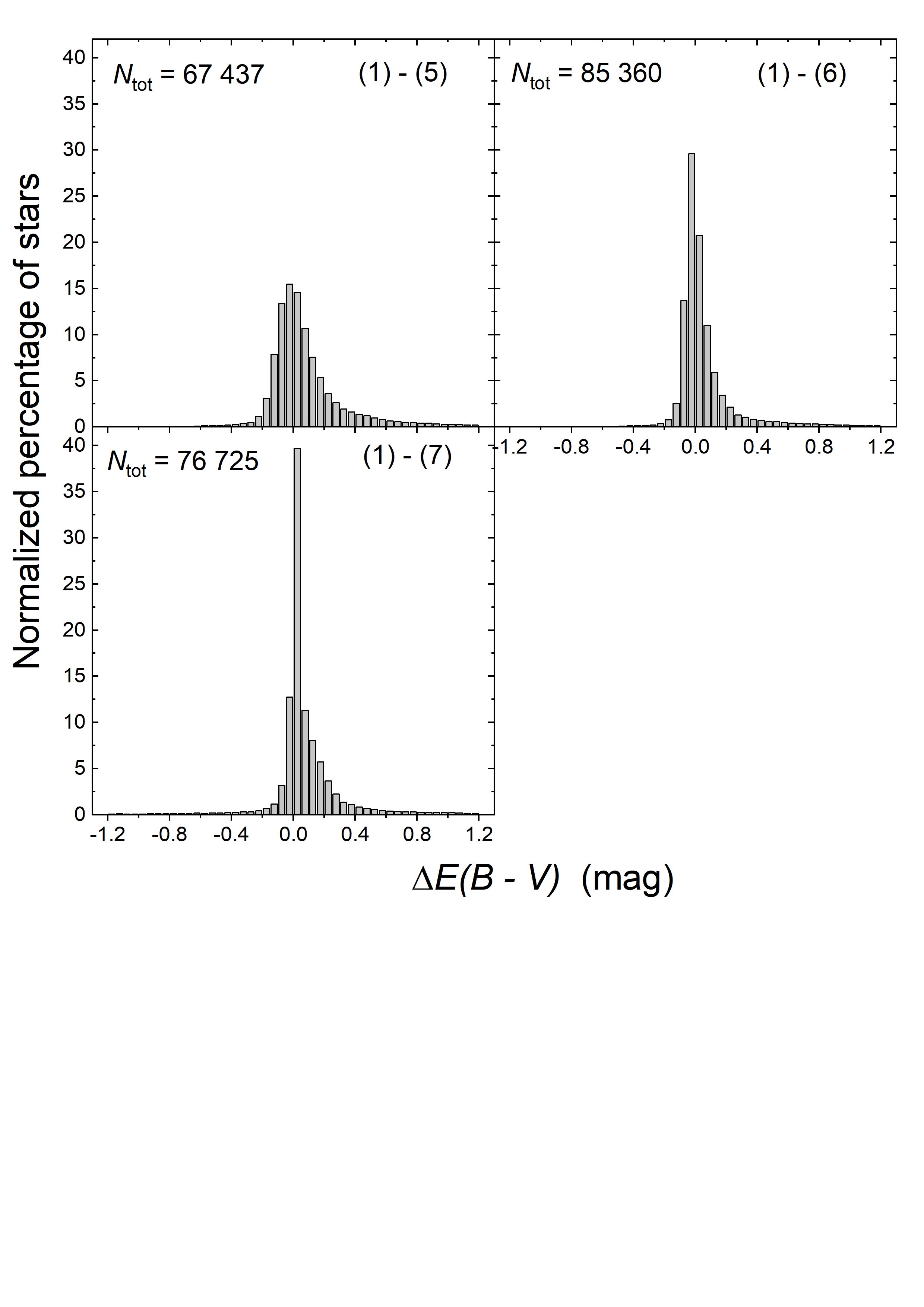}
\caption{Histograms of the differences for
(1) this work, (2) \citet{2019MNRAS.483.4277C}, (3) \textit{Gaia} DR2 \citet{2018A&A...616A...8A}, (4) \textit{Gaia} DR3 \citep{2021A&A...649A...1G}, (5) StarHorse \citep{2019A&A...628A..94A},
(6) StarHorse2021 \citep{2022A&A...658A..91A}, and  (7) Bayestar2019 \citep{2019ApJ...887...93G},
as listed in Table \ref{statistics_published_catalogues}.} 
\label{differences_4} 
\end{center} 
\end{figure*}

\end{appendix}

\end{document}